\documentclass{llncs}
\usepackage{graphicx}
\usepackage{color}
\usepackage{todonotes}

\newcommand{\nc}{\color{black}}

\usepackage{amsthm}
\usepackage{graphicx,rotating}
\usepackage{complexity}
\usepackage{tikz}
\usepackage{xspace}
\usepackage{fix-cm}
\usetikzlibrary{shadows}
\usetikzlibrary{shapes,positioning,arrows}
\usetikzlibrary{decorations.pathmorphing}
\usetikzlibrary{shadows,arrows,patterns,automata,calc,intersections,fit,shapes.misc, decorations.markings}
\usepackage[underline=false,rounded corners=true]{pgf-umlsd}
\usepackage{todonotes}
\usepackage{paralist}
\setdefaultenum{1)}{(a)}{i)}{A)}

\usepackage{msc}

\usepackage[
bookmarks=false,
breaklinks=true,
colorlinks=true,
linkcolor=black,
citecolor=black,
urlcolor=black,
pdfpagelayout=SinglePage
]{hyperref}
\usepackage[all]{hypcap}

\usepackage{dsfont}
\usepackage[bottom]{footmisc}
\usepackage[nocompress]{cite}
\usepackage{listings}
\usepackage{fancyvrb}
\usepackage{cleveref}
\usepackage{bm}
\usepackage{amssymb}
\usepackage{amsmath}
\usepackage{mathtools}
\usepackage{etoolbox}
\usepackage{varwidth} 
\usepackage{array} 
\usepackage{chngcntr}
\usepackage{stmaryrd}
\usepackage{tabulary}
\usepackage{booktabs}

\usepackage{wrapfig,epsfig}

\usepackage{xcolor}
\usepackage{microtype}
\usepackage[T1]{fontenc}
\usepackage{textcomp}
\usepackage[utf8]{inputenc}
\usepackage{inconsolata}

\usepackage{enumitem}

\usepackage[linesnumbered, ruled, noend]{algorithm2e} 
\usepackage{algorithmicx}

\usepackage{subfig}
\usepackage{float}
\usepackage{listing}
\usepackage{mdframed}


\lstdefinestyle{multiline}{
basicstyle=\scriptsize\sffamily,
numbers=left,
numberstyle=\tiny,
frame=tb,
columns=fullflexible,
showstringspaces=false,
escapeinside=\`\`
}

\let\bbordermatrix\bordermatrix
\patchcmd{\bbordermatrix}{8.75}{4.75}{}{}
\patchcmd{\bbordermatrix}{\left(}{\left[}{}{}
\patchcmd{\bbordermatrix}{\right)}{\right]}{}{}

\let\abordermatrix\bordermatrix
\patchcmd{\abordermatrix}{8.75}{4.75}{}{}
\patchcmd{\abordermatrix}{\left(}{\left\langle}{}{}
\patchcmd{\abordermatrix}{\right)}{\right\rangle}{}{}
\usepackage{etoolbox}

\AtBeginDocument{%
  \mathchardef\mathcomma\mathcode`\,
  \mathcode`\,="8000
}
{\catcode`,=\active
  \gdef,{\mathcomma\discretionary{}{}{}}
}

\newcolumntype{M}{>{\begin{varwidth}{1.5cm}}l<{\end{varwidth}}} 
\makeatletter
\newcommand{\raisemath}[1]{\mathpalette{\raisem@th{#1}}}
\newcommand{\raisem@th}[3]{\raisebox{#1}{$#2#3$}}
\makeatother

\counterwithin*{equation}{definition}
\theoremstyle{remark}

\theoremstyle{remark}

\numberwithin{mysubcase}{mycase}

\definecolor{codegreen}{rgb}{0,0.6,0}
\definecolor{codegray}{rgb}{0.5,0.5,0.5}
\definecolor{codepurple}{rgb}{0.58,0,0.82}
\definecolor{backcolour}{rgb}{0.95,0.95,0.92}

\definecolor{bluekeywords}{rgb}{0.13,0.13,1}
\definecolor{greencomments}{rgb}{0,0.5,0}
\definecolor{redstrings}{rgb}{0.9,0,0}

\lstdefinestyle{tamarinStyle}{
  showspaces=false,
  showtabs=false,
  breaklines=true,
  showstringspaces=false,
  breakatwhitespace=true,
  escapeinside={(*@}{@*)},
  basicstyle=\footnotesize\ttfamily,
  breakatwhitespace=false,
  breaklines=true,
  captionpos=b,
  keepspaces=true,
  showspaces=false,
  showstringspaces=false,
  showtabs=false,
  tabsize=2
}

\restylefloat{figure}

\crefname{algocf}{alg.}{algs.}
\Crefname{algocf}{Algorithm}{Algorithms}
\crefalias{AlgoLine}{line}
\makeatletter
\let\cref@old@stepcounter\stepcounter
\def\stepcounter#1{%
  \cref@old@stepcounter{#1}%
  \cref@constructprefix{#1}{\cref@result}%
  \@ifundefined{cref@#1@alias}%
    {\def\@tempa{#1}}%
    {\def\@tempa{\csname cref@#1@alias\endcsname}}%
  \protected@edef\cref@currentlabel{%
    [\@tempa][\arabic{#1}][\cref@result]%
    \csname p@#1\endcsname\csname the#1\endcsname}}
\makeatother
\newcommand{\Fref}[1]{Figure~\ref{#1}}

\newcommand{\Tref}[1]{Table~\ref{#1}}

\newcommand{\Sref}[1]{Section~\ref{#1}}

\newcommand{\tamarin}{T\textsc{amarin}\xspace}
\newcommand{\proverif}{P\textsc{ro}V\textsc{erif}\xspace}
\newcommand{\rewire}{R\textsc{ewire}\xspace}
\newcommand{\basic}{P\textsc{lain}\xspace}
\newcommand{\rtoken}{R\textsc{-token}\xspace}
\newcommand{\otoken}{O\textsc{-token}\xspace}
\newcommand{\otokens}{O\textsc{-tokens}\xspace}
\newcommand{\otokenFull}{O\textsc{bscure} T\textsc{oken}\xspace}

\newcommand{\osrreq}{O\textsc{sr-req}\xspace}
\newcommand{\osrconf}{O\textsc{sr-conf}\xspace}

\newcommand{\Vj}{$V_j$}
\newcommand{\VjX}[1]{$V_#1$}
\newcommand{\SkVj}{$SK_{V_j}$}
\newcommand{\SkVx}[1]{$SK_{V_#1}$}
\newcommand{\PkVj}{$PK_{V_j}$}
\newcommand{\PsiVj}{$Ps_{i(V_j)}$}
\newcommand{\PsxVj}[1]{$Ps_{#1(V_j)}$}
\newcommand{\PsxVy}[2]{$Ps_{#1(V_#2)}$}
\newcommand{\SkPsiVj}{$SK_{Ps_i(V_j)}$}
\newcommand{\PkPsiVj}{$PK_{Ps_i(V_j)}$}
\newcommand{\SkRa}{$SK_{RA}$}
\newcommand{\PkRa}{$PK_{RA}$}
\newcommand{\LtkVj}{$LTK_{V_j}$}

\newcommand{\SkOPsiVj}{$SK_{O_{Ps_i(V_j)}}$}
\newcommand{\PkOPsiVj}{$PK_{O_{Ps_i(V_j)}}$}
\newcommand{\rtokenNotation}{$\sigma_{Ps_i(V_j)}$}
\newcommand{\rtokenNotationXY}[2]{$\sigma_{Ps_#1(V_#2)}$}
\newcommand{\otokenNotation}{$\phi_{Ps_i(V_j)}$}
\newcommand{\otokenNotationX}[1]{$\phi_{Ps_#1(V_j)}$}

\newcommand{\specialcell}[2][c]{%
  \begin{tabular}[#1]{@{}l@{}}#2\end{tabular}}
\newcommand{\ra}[1]{\renewcommand{\arraystretch}{#1}}

\definecolor{codegray}{gray}{0.9}

\newcommand{\textbfi}[1]{\textbf{\textit{#1}}} 

\graphicspath{./ {./figures/img/} }
\begin{document}
\title{Formal Analysis of V2X Revocation Protocols}
\titlerunning{\otoken}
%


\author{
Jorden Whitefield\inst{1}
\and Liqun Chen\inst{1}
\and Frank Kargl\inst{2}
\and Andrew Paverd\inst{3}
\and Steve Schneider\inst{1}
\and Helen Treharne\inst{1}
\and Stephan Wesemeyer\inst{1}}
\authorrunning{Jorden Whitefield et al.} 
%
\tocauthor{Jorden Whitefield, Liqun Chen, Frank Kargl, Steve Schneider, Helen Treharne and Stephan Wesemeyer}
\institute{Department of Computer Science, University of Surrey, Guildford, UK\\
\email{\{J.Whitefield, Liqun.Chen, S.Schneider, H.Treharne, S.Wesemeyer\}@surrey.ac.uk}
\and Ulm University, Ulm, Germany\\
\email{frank.kargl@uni-ulm.de}
\and Aalto University, Espoo, Finland\\
\email{andrew.paverd@ieee.org}}


%
\maketitle              
\begin{abstract}

Research on vehicular networking (V2X) security has produced a range of security
mechanisms and protocols tailored for this domain, addressing both security and
privacy. Typically, the security analysis of these proposals has largely been
informal. However, formal analysis can be used to expose flaws and ultimately
provide a higher level of assurance in the protocols.
This paper focusses on the formal analysis of a particular element of security
mechanisms for V2X found in many proposals, that is the revocation of malicious or misbehaving
vehicles from the V2X system by invalidating their credentials. This revocation needs to be performed in an unlinkable way for vehicle privacy even in the
context of vehicles regularly changing their pseudonyms. 
The \rewire scheme by Förster et al. and its subschemes \basic
and \rtoken aim to solve this challenge by means of cryptographic solutions and
trusted hardware.
Formal analysis using the \tamarin prover identifies two flaws:
 one previously reported in the literature concerned with functional correctness of the protocol, and one previously unknown flaw concerning an authentication property of the \rtoken scheme.  In response to these flaws we propose
\otokenFull (\otoken), an extension of \rewire to enable revocation in a privacy
preserving manner. Our approach addresses the functional and authentication
properties by introducing an additional key-pair, which offers a stronger and
verifiable guarantee of successful revocation of vehicles without resolving the
long-term identity. Moreover \otoken is the first V2X revocation protocol to be
co-designed with a formal model.
\end{abstract}

\begin{keywords}
ad hoc networks, authentication, security verification, V2X.
\end{keywords}
\section{Introduction}
\label{sec:intro}
The term Intelligent Transportation Systems (ITS) denotes the on-going trend to include information and communication technologies (ICT) in vehicles and transportation infrastructure in order to enable safer, coordinated, environmentally friendly, and smarter transportation networks~\cite{willke2009survey}.
Having smarter transportation systems typically involves extending the communication capabilities between the involved entities.

This goes by the term ``Vehicle-to-X (V2X)'' communication and involves various forms of ad-hoc and cellular networking among vehicles and traffic infrastructure. Security and privacy in V2X have played an important role right from the start~\cite{papadimitratos2008secure}.

In particular, anonymity is a requirement in a V2X network as various privacy issues arise from the frequent and real-time broadcasting of the position of vehicles in an
ITS~\cite{schaub2009privacy}, as otherwise mobility patterns can easily be identified. This
makes tracking and profiling of entities possible, which can be used to
systematically collect and infer private information. Pseudonym certificates (pseudonyms)
\cite{DBLP:journals/comsur/PetitSFK15} are the most commonly applied way to
address privacy concerns and are also foreseen in emerging standards.

Schaub et al.~\cite{schaub2009privacy} discuss various requirements for such a pseudonym system and
Petit et al.~\cite{DBLP:journals/comsur/PetitSFK15} survey a large body of existing work and from there identify an abstract
pseudonym life cycle which is comprised of five main phases: issuance, use, change,
resolution and revocation. Within an ITS architecture there are three trusted
third parties that support the life cycle of pseudonyms: a certification authority (CA), a provider of pseudonyms (PP), and a
revocation authority (RA). The CA issues
long-term credentials to vehicles. The PP is responsible for handing out shorter-lived pseudonym certificates. The RA receives and collects information such as reports on misbehaviour, takes decisions to revoke a misbehaving entity, and implement this revocation by whatever means a specific scheme foresees.

Effective revocation has been identified as a
challenge~\cite{Raya:2007:EMF:2312265.2316074} due to the decentralised nature
of vehicle networks and the ability of vehicles to change their active pseudonyms.

{\it Related work.}
Pseudonym revocation techniques have largely been based on the distribution
of certificate revocation lists (CRLs)~\cite{Raya:2007:EMF:2312265.2316074,DBLP:journals/comsur/PetitSFK15},
such that when a misbehaving vehicle is revoked an updated CRL is broadcast to all
vehicles. Several approaches have been taken to optimise the
protocols and distribution process of CRL delivery~\cite{SRAAC,papadimitratos2008certificate,
ma2008pseudonym,michael2010scalable,kondareddy2010analysis,haas2011efficient}.
However, these approaches often either revoke only one pseudonym of a vehicle -- thereby missing the goal of removing a misbehaving vehicle completely -- or they create a way of linking pseudonyms -- then hurting privacy protection.

Bi{\ss}meyer et al.~\cite{bissmeyer2013copra} propose the CORPA protocol that
allows conditional
pseudonym resolution which preserves the privacy.

Raya et al. propose an infrastructure-based revocation protocol ~\cite{toor2008vehicle}, which
remotely deletes keys in a trusted component. Their protocol requires that a
vehicle's identity is known to perform revocation, in combination with a CRL -- again a clear drawback with respect to privacy.

Schaub et al. propose $V$-Tokens~\cite{DBLP:conf/wcnc/SchaubKMW10}, which introduces
embedding vehicle resolution information directly into pseudonyms. A V-Token
is a ciphertext field in the pseudonym certificate that is created from a vehicle's identity,
the CA's identity and a randomisation factor $r$ all encrypted with the RA's public key.
In this scheme multiple trusted parties need to collaborate to resolve the pseudonym, which then reveals
a vehicle's identity that is used for revocation. In case of a revocation, this therefore violates the privacy
of vehicles, as resolution of their pseudonym to an identity is required.

F{\"o}rster et al. propose PUCA~\cite{Forster2016122}, a pseudonym scheme based on anonymous credentials where privacy of the vehicle owner
has absolute priority and no way exists for resolving pseudonyms. PUCA foresees no way of credential revocation.
However, the same authors then also propose \rewire~\cite{DBLP:conf/trust/ForsterLZK15}, a
modular revocation mechanism within a decentralised network which is not relying on
the resolution approach that can be used to introduce revocation in PUCA. Instead, \rewire assumes on-board Trusted
Components (TC) in vehicles to support revocation.

A series of EU research projects, e.g.,  SeVeCom~\cite{papadimitratos2007architecture},
EVITA~\cite{EVITA:HRSW09} and PRECIOSA~\cite{PRECIOSA} have investigated securing
V2X architectures using TCs. The recent project PRESERVE~\cite{PRESERVE:website}
has even prototyped this in an ASIC for secure ITS.
Feiri et al.~\cite{DBLP:conf/vnc/FeiriPK14} propose to use
TCs to store pseudonyms in secure storage and use a physical-unclonable function (PUF) to reduce the need for large amounts of secure storage.
Based on such earlier work, it seems a reasonable assumption that hardware security modules (HSMs) are available as trust anchors, as done in
the specification of our \otoken approach.

In this paper we explore the two versions of the \rewire protocols~\cite{DBLP:conf/trust/ForsterLZK15}, which are referred to as \rewire \basic
and \rtoken. This protocol represents the current state of the art of those proposed for revocation in V2X architectures. No revocation protocol has been deployed in vehicles as yet.

{\it Contribution.} In this paper, we describe the formalisation of the revocation protocols
proposed by F{\"o}rster et al.~\cite{DBLP:conf/trust/ForsterLZK15}, which was done using multiset rewriting as
supported by the \tamarin prover. These protocols have not previously been formally
analysed. 
We present definitions of functional correctness and authentication as properties of the protocols.
Our formal analysis reveals that the \basic model does not preserve functional correctness, specifically that
a vehicle is not guaranteed to be revoked and therefore could
continue to participate in communication messages within an ITS. 
This formally confirms a flaw that was observed by F{\"o}rster et al.~\cite{DBLP:conf/trust/ForsterLZK15}. Our
analysis of the \rtoken protocol identifies a hitherto unknown flaw: that it does not guarantee authentication
properties, in particular it does not guarantee that the confirmation of revocation actually came from the
intended vehicle. This new unknown weakness is acknowledged by the authors of the \rtoken protocol as a flaw.

The insights gained from the formal modelling motivated our proposal for a new
protocol. We therefore develop a new protocol which proposes improvements to the
\rewire protocols that ensures correct revocation of an entity under any
pseudonym without requiring resolution even if its active pseudonym has changed
by the time of revocation. In this paper we refer to our new
protocol as the \otokenFull (\otoken) protocol. Its novelty is the inclusion of
an additional asymmetric key pair for signature, used to augment the pseudonyms that are utilised within
message exchanges for verifiable revocation. The new protocol is shown to preserve
all the desired authentication and functional correctness properties.
 Our proposed protocol, similar to the previous protocols discussed in this paper, 
  requires a trusted device at the car which will engage in the
  revocation protocol and on completion can be trusted to erase all of
  the pseudonyms that the car may have available. 

Due to limited space, we will not present the details of the \tamarin model rules and lemmas in this paper.  The models of the three protocols presented in this paper have been made available~\cite{REWIRE:Models}. 

{\it Structure.} This paper is organised as follows:~\Sref{sec:scenario} presents
a revocation scenario. \Sref{sec:tands} introduces \tamarin, together with
 the security notation used throughout the paper and the modelling
assumptions made in our symbolic models. \Sref{sec:rewire} defines formal models
and evaluates the existing \rewire protocols. \Sref{sec:otoken} presents our
new enhanced \rewire protocol and its analysis and~\Sref{sec:conc} finally provides
conclusions and identifies preservation of privacy properties as an area of
future analysis for revocation protocols.
\section{System Model and Revocation Scenario}
\label{sec:scenario}
The process of revocation for the existing \rewire protocols and
our \otoken protocol follows the same pattern shown in~\Fref{fig:img:RevScen}.
\Fref{fig:img:RevScen} illustrates the three main authorities in an ITS, namely
the CA, the PP and the RA, and how vehicles interact with them.
The purpose of these authorities and vehicles in a revocation scenario is as follows:

\begin{figure}[t!]
\vspace{-15pt}
\centering
\includegraphics[width=0.95\textwidth]{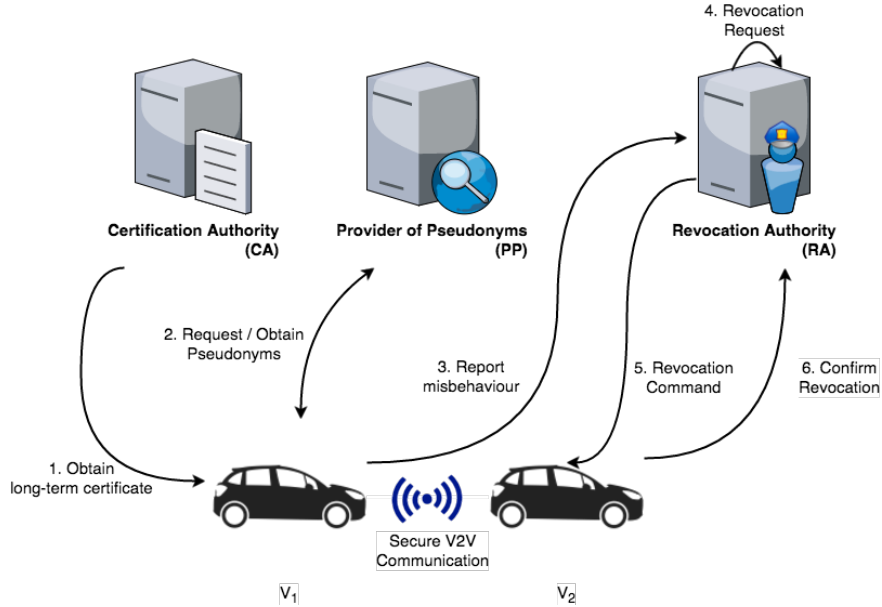}
\caption{High-level V2X Revocation Scenario}
\label{fig:img:RevScen}
\vspace{-5pt}
\end{figure}

{\let\labelitemi\labelitemii
\begin{itemize}
\item The CA and PP issue long-term certificates and pseudonyms respectively to
vehicles and may optionally implement a resolution mechanism to allow linking back
pseudonyms to long-term IDs.

\item Vehicles in the ITS communicate with other participants. They monitor each
others behaviour using misbehaviour detection mechanisms~\cite{DBLP:journals/corr/HeijdenDLK16}
and may issue reports of vehicle misbehaviour to the RA.

\item The RA collects misbehaviour reports from participating vehicles
in an ITS, and takes a decision to revoke reported pseudonyms. It
then creates and broadcasts signed revocation messages to the misbehaving
vehicle.

\item Vehicles receive and process revocation commands to revoke their pseudonyms,
and send confirmations back to the RA.
\end{itemize} }

The \rewire protocols and our variant has the following steps: In step
1 vehicle \VjX{1} obtains a long-term certificate from the
CA enabling it to obtain pseudonyms. In step 2 \VjX{1} obtains pseudonyms from the
PP to communicate securely with other vehicles including vehicle \VjX{2}.
Steps 1 and 2 are not part of a revocation protocol itself, rather they are part of
the issuance phase of pseudonyms. During the communication in the ITS, vehicle
\VjX{1} will receive messages from \VjX{2} under a pseudonym which could be changed
frequently. \VjX{2} will apply misbehaviour detection
mechanisms~\cite{DBLP:journals/corr/HeijdenDLK16} in
order to detect indications of faulty or malicious behaviour. Examples of such
mechanisms may detect spoofed positions or incorrect speeds reported in messages.

In such cases, step 3 is triggered by \VjX{1} submitting a misbehaviour report to
the RA accusing \VjX{2} of misbehaviour. Similarly other vehicles may make the same
report to the RA against \VjX{2} (omitted from the diagram). The RA takes a decision
to have \VjX{2}'s access to the ITS infrastructure revoked if some threshold is reached.
Then the RA crafts a report containing the reason for revocation
and \VjX{2}'s current pseudonym (step 4). Following the receipt, a revocation message
is broadcast to all vehicles in step 5. \VjX{2} receives the designated revocation
message and its TC will be triggered to delete all of its pseudonyms. Finally,
\VjX{2} constructs and sends a confirmation message back to the RA in step 6 to
inform the RA all of its pseudonyms were deleted.
\section{Background and assumptions}
\label{sec:tands}
\subsection{\tamarin}
\label{sec:rp:tamarin}
We model and analyse all three protocols, the \basic and \rtoken protocols
and our new \otoken protocol in~\Sref{sec:otoken} using the \tamarin prover.  For this paper we give a general description of what the \tamarin tool provides.  There are several full introductions to the tool~\cite{TamarinWebsite,DBLP:conf/csfw/SchmidtMCB12,DBLP:conf/cav/MeierSCB13} for further details.

The \tamarin prover is a symbolic analysis tool that is based on multi-set rewriting rules and first order logic.  It supports the analysis of security protocols, which are described using multi-set rewrite rules to describe actions corresponding to protocol agents taking part in protocol steps.  Protocol messages are modelled as terms which enable cryptographic protocol constructions such as encryption, decryption, signatures and hashing to be expressed.  Thus the terms in \Tref{tab:tands:notation} are all expressible in \tamarin syntax.  The global state of the system is captured as a multi-set of Facts, which are expressed as predicates on terms, of the form $F(t_1,\ldots,t_n)$.  A rewrite rule, labelled by an action, takes a multi-set of facts, and replaces (or rewrites) them with another multi-set of facts, labelled by an action.

A Dolev-Yao adversary is also built into the tool.  The rewrite rules induce a transition system describing the potential executions of (unbounded numbers of) protocol instances in the context of the adversary.  The transition system has a formal semantics which underpins the soundness of the tool.

Properties on the actions can be expressed using first-order logic, enabling requirements on executions to be defined.  \tamarin enables the analysis of the transition system with respect to such properties.  Authentication properties are typically of the form ``for every execution, if action $a_2$ occurs then action $a_1$ must previously have occurred''.  For example, if $a_2$ corresponds to agent $A$ receiving a confirmation message, and $a_1$ corresponds to $B$ sending that message, then the authentication property is that $A$'s receipt of the message guarantees that $B$ sent it (i.e., it was not spoofed by the adversary).  If every execution satisfies this property then the protocol provides the authentication required.

\tamarin has numerous built-in security theories that abstractly support common
cryptographic functions. For example, in this paper we use the {\it signing}
built-in which models a signature scheme. It provides the function symbols
\texttt{sign} and \texttt{verify} such that digital signatures can be verified
using the equation: \\ \texttt{verify( sign( m, sk ), m, pk( sk ) ) = true}.

%
%
\subsection{Security Notation and Analysis}
\label{sec:tands:notation}
The notation defined in~\Tref{tab:tands:notation} is used across
all models in the paper. The last three entries are specific to our new
protocol in~\Sref{sec:otoken}.
%
\begin{table*}[t!]\centering
\vspace{-5pt}
\ra{1.2}
    \begin{tabular}{llll}
        \toprule
        {\bf Syntax} & & & {\bf Description}\\
        \midrule
        \Vj & & & \specialcell{~An arbitrary vehicle $j$}\\
        \SkVj & \PkVj & & \specialcell{~Asymmetric key pair for \Vj }\\
        \PsiVj & & & ~$i^{th}$ pseudonym of \Vj \\
        \SkPsiVj~ & \PkPsiVj & & \specialcell{~Asymmetric pseudonym
                                                key pair for \Vj's~$i^{th}$ \\~pseudonym}\\
        \SkRa & \PkRa & & \specialcell{~Asymmetric key pair for the RA}\\
        \multicolumn{3}{l}{\rtokenNotation$~:=~\{|~$\Vj$~||~$\PkVj$~||~r~|\}_{SK_{V_j}} $} & \specialcell{~An \rtoken of the $i^{th}$ pseudonym of $V_j$, \\~where $r$ is a nonce} \\
        \hline
        \LtkVj & & & ~\specialcell{Long-term symmetric key of a vehicle \Vj\\(replaces asymmetric pair in line 2 above)}\\
        \SkOPsiVj~ & \PkOPsiVj & & \specialcell{~Asymmetric key
                                                      pair for an \otoken, \\~belonging to the
                                                      $i^{th}$ pseudonym of \Vj}\\
        \multicolumn{3}{l}{\otokenNotation$~:=~\{|~$\SkPsiVj$~|\}_{LTK_{V_j}}$} & \specialcell{~An \otoken of the $i^{th}$ pseudonym of $V_j$}\\
        \bottomrule
    \end{tabular}
\caption{Security Notation}
\label{tab:tands:notation}
\vspace{-25pt}
\end{table*}
The following seven proof goals are considered in this paper to model our correctness requirements.

{\it G1: Executable} ensures the model is executable and demonstrates successful transmission of all core messages.  It is a sanity check of the model's correctness.

{\it G2: Weak agreement}, defined by Lowe~\cite{Lowe}, is a form of authentication which guarantees that when an
initiator $A$ completes a run of the protocol then it was interacting
with another agent $B$ who had also been running the protocol. In the revocation
protocols the initiator $A$ is the RA and an agent $B$ is a vehicle.

{\it G3: Non-injective agreement}, again defined by Lowe~\cite{Lowe},  adds a further condition to ensure that the two agents, $A$ and $B$, agree on the roles they are taking and agree on the data items
used in their message exchange. In our protocols {\it non-injective agreement}
guarantees that the RA and vehicle both agree upon the completion of a run with
each other and that in those runs the contents of the received messages correspond
to the sent messages.

{\it G4: Non-injective synchronisation}, defined by Cremers and Mauw~\cite{DBLP:books/sp/Cremers12}, is very similar to {\it non-injective agreement}
but additionally requires that the corresponding send and receive messages have
to be executed in their expected order. This means that in the revocation protocols
revoke messages are sent later than receive messages. This means that if a
protocol preserves a {\it non-injective synchronisation} property then the
corresponding {\it non-injective agreement} property will also hold.

{\it G5: Revoke after change esists}, defined in this paper, states that if a vehicle changes its pseudonym and a previous pseudonym is revoked, it should still be possible for the vehicle to create a message to confirm the Revocation Authority (RA) that it has taken the action for revocation.  This is a sanity check that the a vehicle can be revoked even after a change of pseudonym.

{\it G6: Order for self revocation (OSR) request received with change all}, defined in this paper, indicates that if a vehicle receives the OSR request, the vehicle will perform the revocation and create a confirmation.

{\it G7: Revoke with change all}, defined in this paper, states that if a confirmation of a pseudonym revocation is accepted by the RA from a vehicle then that vehicle will have accepted and processed a revocation request from the RA.

\nc



%
\subsection{Modelling Assumptions}
\label{sec:tands:mass}
In this section we provide a scope for the protocols and identify the modelling
abstractions that are used for the analysis. We assume that for each of the
protocol models a registration and enrolment phase has executed, resulting in
vehicles holding valid pseudonyms. All vehicles in a network have a Trusted
Component (TC) and abstractly this means that 1). vehicle keys cannot be leaked,
and 2). vehicles cannot ignore revocation messages. We consider the CA, PP and
RA to be distinct roles and in the architecture there
is one of each. These roles are all trustworthy and therefore, we remove the
possibility of their keys leaking from the analysis.

Step 1 and 2 in~\Sref{sec:scenario} denotes the issuing of pseudonyms to vehicles by
the CA and will be abstractly captured as a rule within our models. A revocation
protocol focuses on steps 3, 4, 5 and 6 from~\Fref{fig:img:RevScen}.
Within the \tamarin model steps 3 and 4 are abstractly represented by a report
event which the RA receives. Steps 5 and 6 are described in three rules which
focus on the message exchange to revoke a vehicle and a confirmation to affirm
the vehicle followed the request. All the formal models in this paper follow this
pattern of communication but the format of the messages and the verification that
can be performed on the signed messages changes with each protocol.

The Dolev-Yao adversary in our models is in control of the network and other
untrusted parts of the system including the vehicles themselves. It is not in
control of the TCs of the vehicles and the trusted third parties.



\section{\rewire Protocols}
\label{sec:rewire}
This section describes our modelling and analysis of the \basic and \rtoken protocols. Our security
and functional correctness analysis shows the following \textbfi{main results} which are weaknesses
in the existing protocols:
{\let\labelitemi\labelitemii
\begin{itemize}
  \item If the \basic protocol executes a change of pseudonym, then no confirmation guarantee can be
        communicated to the RA. Hence even though authentication
        properties may hold, a misbehaving vehicle may avoid revocation by changing
        its pseudonym, and so functional correctness will not be guaranteed. While the
        original paper~\cite{DBLP:conf/trust/ForsterLZK15} already
		identified this issue and addressed it in the \rtoken version, \tamarin
        was independently able to discover this problem.
  \item Following attempted revocation of a vehicle's pseudonym the RA is unable to verify
        successful confirmation in the \rtoken scheme, thus none of
        the authentication properties hold. In particular a confirmation can be
        spoofed by a malicious agent and accepted by the RA, even when the misbehaving
        vehicle is not revoked. This flaw was not previously recognised.
\end{itemize} }

%
\subsection{\rewire: \basic}
\label{sec:rp:basic}
\textbfi{Modelling.}
\Sref{sec:tands:mass} informally identified the steps of a revocation protocol
based on the behaviour of an RA and a misbehaving vehicle. We model the protocol
roles of the RA and an arbitrary vehicle (\Vj) in \tamarin by a set of rewrite rules, which correspond to the steps of the protocol.
The \basic model has three distinct types of rules to: 1). setup all required
key pairs for secure communication, 2). create misbehaviour reports and 3). describe
revocation requests and receiving subsequent confirmation.

%
The heart of the protocol involves an exchange of messages to effect revocation: an Order for Self-Revocation (OSR) request, followed by a confirmation response.

The OSR request message \osrreq~\cite{DBLP:conf/trust/ForsterLZK15} is the first
message sent to a vehicle, which triggers its revocation process. \osrreq contains
the command to revoke, the reported misbehaving pseudonym and additional
information as to why the revocation occurred. The pseudonym \PsiVj in
this protocol is simply \PkPsiVj belonging to \Vj. \osrreq is signed
by the RA, and can be verified by receiving vehicles.
  \[\osrreq := \{|~``revoke"~||~{Ps_{i(V_j)}}~||~reason~|\}_{SK_{RA}}\]
The \osrreq message is received and verified by a \Vj, and the TC in \Vj
can identify the pseudonym as belonging to \Vj. Following this identification
the vehicle constructs an \osrconf message confirming the command to revoke was
followed, and the TC in \Vj will flag all available pseudonyms as revoked to
prevent their future use in V2X communication. The \osrconf message is comprised
of two terms: a confirm command and the active reported public pseudonym key.

The message is signed with the corresponding secret pseudonym key.
  \[\osrconf := \{|~``confirm"~||~{Ps_{i(V_j)}}~ |\}_{SK_{Ps_{i(V_j)}}}\]
  
 We model a well formed \osrreq message duly signed
by the RA and addressed to its current pseudonym. The vehicle
verifies that the message came from the RA and contains the vehicle's active pseudonym,
before deleting all its pseudonyms and creating the \osrconf message signed under
the active secret pseudonym key, which is sent back to the RA. The adversary is
able to learn the \osrreq message terms and the signature. However the adversary
cannot modify the contents of the message as the adversary does not posses the RA's
secret key. 
We also model the incoming \osrconf message
from a $V_j$. The RA verifies the \osrconf message is signed with the
reported pseudonym \PkPsiVj. 

\begin{table*}[t!]\centering
\vspace{-25pt}
\ra{1.2}
    \begin{tabular}{llllclclc}
        \toprule
  {\bf Goal} & & {\bf Content} & & {\bf \basic} & & {\bf \rtoken} & & {\bf \otoken}\\ 
        \midrule
        G1 & & \texttt{executable} & & \checkmark & & \checkmark & & \checkmark \\
        G2 & & \texttt{weak\_agreement} & & \checkmark & & $\times$ & & \checkmark \\
        G3 & & \texttt{noninjective\_agreement} & & \checkmark & & $\times$ & & \checkmark \\
        G4 & & \texttt{noninjective\_synchronisation} & & \checkmark & & $\times$ & & \checkmark \\
        G5 & & \texttt{revoke\_after\_change\_exists} & & $\times$ & & \checkmark & & \checkmark \\
        G6 & & \texttt{osr\_req\_received\_with\_change\_all} & & n/a & & \checkmark & & \checkmark \\
        G7 & & \texttt{revoke\_with\_change\_all} & & n/a & & $\times$ & & \checkmark \\
        \bottomrule
    \end{tabular}
\caption{Summary of results}
\label{tab:tands:results}
\vspace{-5pt}
\end{table*}

\textbfi{Proof Goals.} We state several proof goals for our model, G1-G7 discussed in~\Sref{sec:tands:notation},  that represent
authentication and functional correctness properties. The results of whether each
of the numbered proof goals hold are summarised in~\Tref{tab:tands:results}.
All the goals include predicates requiring that the vehicle's long-term key and
secret pseudonym keys are not compromised, and so correct behaviour is dependent on these keys not being compromised. 


%
A successful run of the model guarantees that \Vj was running the protocol with the RA.
Receipt of the \osrconf message represents completion of a run for the RA.  An \osrreq
message is represented by facts from both the RA and vehicle's perspective. 
The model observes that the RA will have completed a run
and verified a confirmation from a vehicle. Furthermore, the vehicle must have
received the \osrreq message before it is possible for the RA to receive the
\osrconf message, hence the communication order is preserved.
The above proof goals are trace authentication properties demonstrating that the
attacker cannot construct \osrreq or \osrconf messages from its observations.
Thus no logical attacks are identified for the \basic protocol from
our symbolic analysis.


\subsection{\rewire: \basic with change of pseudonym}
\label{sec:rp:basic_with_change}
\begin{figure}[t!]
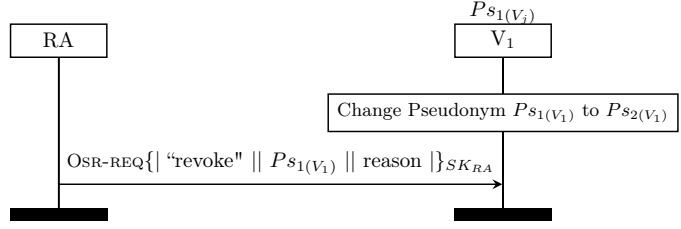

\vspace{-45pt}
\centering
\scalebox{.8}{
\setmsckeyword{}
    \begin{msc}{}
        \drawframe{no}
        \setlength{\instdist}{9.6\instdist}

        \declinst{RA}{}{RA}
        \declinst{V}{\PsxVj{1}}{V$_1$}

        \action*{\small Change Pseudonym \PsxVy{1}{1} to \PsxVy{2}{1}}{V}
        \nextlevel[3]
        \mess{{\small \osrreq}\{| ``revoke" || \PsxVy{1}{1} || reason |\}$_{SK_{RA}}$}{RA}{V}

    \end{msc}}
\caption{\rewire: \basic Pseudonym Scheme incomplete run}
\label{fig:msc:bps}
\vspace{-15pt}
\end{figure}

\textbfi{Modelling.}
In the \basic pseudonym scheme revocation of \rewire~\cite{DBLP:conf/trust/ForsterLZK15},
a {\it change of pseudonym} for a vehicle can occur at any point prior to an \osrreq being
received. For example, consider a vehicle \VjX{1} and two of its pseudonyms \PsxVy{1}{1}
and \PsxVy{2}{1} in the following change of pseudonym scenario. When the RA
receives a report to revoke \VjX{1}, it broadcasts the \osrreq
message containing the misbehaving pseudonym \PsxVy{1}{1}, as shown in~\Fref{fig:msc:bps}.
However, before an \osrreq message is ever received by
\VjX{1} a change of pseudonym can occur resulting in a new pseudonym now being
active. In an na\"ive implementation, changing to \PsxVy{2}{1} means that the receipt of the \osrreq will be
ignored as the vehicle has deleted its previous pseudonym. Therefore, no \osrconf
message will be generated by \VjX{1} as the vehicle has deleted its previous pseudonyms and the revocation process will fail. Consequently \VjX{1} can
continue to misbehave under the new pseudonym \PsxVy{2}{1}.

We model the changing of pseudonyms in such a way that the model creates a fresh pseudonym key for an arbitrary vehicle \Vj. The
``can change'' fact is included to control when a vehicle can change its
current pseudonym. The model concludes by storing the new pseudonym secret key for
\Vj and outputs the public key of the new pseudonym, which the intruder learns.

\textbfi{Proof Goals.}  Adding this extra behaviour to the protocol yields another proof goal, G5,  discussed in~\Sref{sec:tands:notation}.
If a vehicle changes its pseudonym and a previous pseudonym
is revoked, it should be possible for the vehicle to create an \osrconf message.
This model fails for the \basic protocol, showing that the protocol does not
guarantee a successful revocation of a misbehaving vehicle in the presence of
changing pseudonyms, and indeed that if a vehicle changes its pseudonym then it can escape revocation. Therefore, the \basic protocol is not functionally correct in the context of changing pseudonyms.
To address this shortcoming in~\cite{Forster2016122,DBLP:conf/trust/ForsterLZK15}
a variant to the \basic protocol is proposed, referred to as the \rtoken protocol.


\subsection{\rewire: \rtoken}
\label{sec:rp:rtoken}
\textbfi{Modelling.}
The \rtoken variant embeds additional information in pseudonym certificates with
the aim of allowing revocation even with changing
pseudonyms. This additional information is an \rtoken, \rtokenNotation, which is
constructed from a vehicle's public identity, public key and a nonce $r$,
encrypted under a vehicle's secret key. There is a fresh \rtoken for each pseudonym.
\PsiVj in this protocol is a pseudonym containing \PkPsiVj and
the \rtoken \rtokenNotation. 

It is the purpose of the \rtoken to allow a vehicle to later detect whether a revocation request is directed to it, without allowing others to identify the vehicle.
By encrypting the \rtoken under \SkVj, all vehicles must attempt to decrypt
the \rtoken. Only the correct vehicle can decrypt the \rtoken, meaning the revocation was
designated for the vehicle and should be executed.

In PUCA and \rewire
a ``cut and choose" approach~\cite{rabin1979digitalized} is used to generate the
\rtoken, but in the model we have simply abstracted this to a fresh value that is
encrypted under the secret key of the vehicle.

\begin{figure}[t!]
\vspace{-55pt}
\centering
\scalebox{.8}{
\setmsckeyword{}
    \begin{msc}{}
        \drawframe{no}
        \setlength{\instdist}{9.6\instdist}

        \declinst{RA}{}{RA}
        \declinst{V}{}{\VjX{1}}

        \action*{\small Change \PsxVy{1}{1} to \PsxVy{2}{1}}{V}
        \nextlevel[3]

        \mess{{\small \osrreq}\{|~``revoke"~||~\PsxVy{1}{1}~||~reason~|\}$_{SK_{RA}}$}{RA}{V}
        \nextlevel[1]

        \action*{\small Decrypt \rtokenNotationXY{1}{1} with \SkVx{1}} {V}
        \nextlevel[4]

        \mess{{\small \osrconf}\{| ``confirm" || \rtokenNotationXY{1}{1} |\}$_{SK_{V_1}}$}{V}{RA}
        \nextlevel[1]

        \ncondition*{{\it ni-sync}}{RA}
        \nextlevel[1]

    \end{msc}}
\caption{\rewire: \rtoken Scheme}
\label{fig:msc:r}
\vspace{-25pt}
\end{figure}

The \rtoken protocol is represented in~\Fref{fig:msc:r}.
The \osrreq message is of the same format as the \basic protocol where the pseudonym
contains the \rtoken. Once a vehicle receives an \osrreq it attempts to decrypt
the \rtoken irrespective of its active
pseudonym. Only the designated vehicle can decrypt the \rtoken since the decryption
uses \SkVj, others will simply ignore the \osrreq. The \osrconf message now contains
the \rtoken and not the pseudonym, and the message is signed with the vehicle's
secret key.
  \[\osrconf := \{|~``confirm"~||~\sigma_{Ps_i(V_j)}~|\}_{SK_{V_j}}\]
  
The modelling of the rules for the \rtoken protocol is almost identical but there
are two important changes. Firstly the model includes having
to decrypt the \rtoken as an additional action.
Secondly, the model is weakened to remove the
\texttt{verify} step (which checks the correctness of the confirmation \rtokenNotation) since the RA is not in possession of the \PkVj.

\textbfi{Proof Goals.} For consistency we analysed functional correctness. All the proof goals for the \basic protocol remain applicable.  Proof goal G5 holds because
any vehicle can create a confirmation message. Two additional goals are included to
analyse the correct behaviour of the vehicle (G6) and RA (G7) in the context of changing
pseudonyms, as shown in~\Tref{tab:tands:results}.  For
each goal we again assume that \SkVj is not compromised. The security
analysis yields that neither of the authentication properties
hold. The adversary is able to intercept the \osrreq message and create a
\osrconf message containing the inferred \rtoken. The adversary then generates a
fresh secret key which is used to sign the \osrconf message. The created
\osrconf is sent to the RA. The RA accepts the confirmation but cannot verify its authenticity
because the \LtkVj is only known to \Vj and CA. Therefore, The RA does not obtain a guarantee that it is communicating with a
running vehicle.  

This flaw in the protocol was not previously recognised, and has been accepted by the designers of the \rtoken protocol.

\section{\otoken Protocol}
\label{sec:otoken}
%
\textbfi{Modelling.} 
To solve the issue of the RA not being able to verify the confirmation message, \osrconf,
we propose the \otoken protocol. Note that the \otoken mimics the \rtoken closely:
the reason for generating different \otokens for each pseudonym is the same as for
the \rtoken, to ensure unlinkability of the vehicle in question. If the \rtoken
or \otoken remained the same, it would act as a vehicle identifier.

We replace the \rtoken in the previous scheme with a simpler construction: an \otoken for the $i^{th}$ pseudonym of \Vj,
\otokenNotation, consisting of an \SkOPsiVj key which is encrypted under
\LtkVj. Each \otoken is fresh and associated with one and only one \PsiVj
pseudonym.
    \[\phi_{Ps_i(V_j)}~:=~\{|~SK_{O_{Ps_i(V_j)}}~|\}_{LTK_{V_j}}\]
The aim of using fresh \SkOPsiVj keys is to make pseudonyms
unlinkable.

The pseudonym also contains one additional field, \PkOPsiVj, which is
the corresponding public key for the particular \otoken.
Therefore, the pseudonym
contains enough information for the RA to verify a received \osrconf message and
for the vehicle to change its pseudonym. 

\begin{figure}[t!]
\vspace{-45pt}
\centering
\scalebox{.75}{
\setmsckeyword{}
    \begin{msc}{}
        \drawframe{no}
        \setlength{\instdist}{9.6\instdist}

        \declinst{ra}{{\small \SkRa, \PkOPsiVj, \otokenNotation}}{RA}
        \declinst{v}{\PkRa, \LtkVj, \SkOPsiVj}{\Vj}

        \action*{\small Change Pseudonym \PsxVj{1} to \PsxVj{2}}{v}{v}
        \nextlevel[3]

        \mess{{\small \osrreq}\{| ``revoke" || \PsxVj{1} || reason |\}$_{SK_{RA}}$}{ra}{v}
        \nextlevel[1]

        \action*{\small Verify and extract \otokenNotationX{1}}{v}{v}
        \nextlevel[2]

        \action*{\small Decrypt \otokenNotationX{1} OR Fail}{v}{v}
        \nextlevel[2]

        \action*{\small Delete all Pseudonyms}{v}{v}
        \nextlevel[4]

        \mess{{\small \osrconf}\{| ``confirm" || \otokenNotationX{1} |\}$_{SK_{O_{Ps_1(V_j)}}}$}{v}{ra}
        \nextlevel[1]

        \action*{\small Verify {\small \osrconf}}{ra}{ra}
        \nextlevel[2]

        \condition*{{\it ni-sync}}{ra}
        \condition*{{\it ni-sync}}{v}
        \nextlevel[1]

    \end{msc}}
\caption{\otoken Revocation}
\label{fig:msc:otokenRevocation}
\end{figure}


A revocation run which uses \otoken is shown in~\Fref{fig:msc:otokenRevocation}. The
\osrreq message is of the same format as the other protocols but the pseudonym
contains the \otoken. The \osrconf message now contains the \otoken and the
message is signed with \SkOPsiVj instead of signing with \LtkVj
which the vehicle extracted earlier:
  \[\osrconf := \{| ``confirm"~||~\phi_{Ps_i(V_j)}~ |\}_{SK_{O_{Ps_i(V_j)}}}\]
The 
subtle 
change in signing the \osrconf message, together
with the RA's knowledge of \PkOPsiVj enables the RA to verify the
confirmation message. 

%
The modelling of the other rules for the \otoken protocol is largely similar but there
are two further changes. Firstly, the rule for receiving the OSR request includes having
to decrypt the \otoken as an additional action, 
Secondly, changing pseudonym behaviour is supported with a new rule, by creating a fresh pseudonym
secret key, a fresh \SkOPsiVj and the newly encrypted \otoken.

\textbfi{Proof Goals.} The results for the formal analysis for the \otoken protocol
is presented in~\Tref{tab:tands:results} and achieves all desired guarantees.
Notably all the authentication properties hold which means that the RA is communicating
with the revoked vehicle and can verify the received confirmation, which was not the case with the
\rtoken protocol. Therefore, all the desired functional correctness properties hold.

\section{Conclusions and Future Work}
\label{sec:conc}
The new \otoken protocol proposed in this paper allows revocation even if vehicles
have  changed pseudonym. It also allows the RA to
verify a confirmation sent by a vehicle that it has deleted its pseudonyms. The
formal analysis establishes that verifying such a confirmation provides a guarantee
that the revocation occurred. We have therefore shown through formal analysis
that the desired functional correctness and authentication properties hold. The
new \otoken protocol for \rewire was developed by first formally modelling and
analysing the two previous variants of \rewire, then identifying weaknesses in
their functional correctness and a failure to meet required authentication properties.

In an implementation of a revocation protocol, heartbeats provide protection against
non-delivery of revocation requests by incorporating such requests within the heartbeats. TCs within a vehicle expect heartbeats (which may contain revocation requests), which are generated by the RA.  TCs will take appropriate action if they are not received, under the assumption that they have been blocked. Therefore, augmenting
a formal analysis with heartbeats will require a more detailed model of a TC and
further adversarial behaviour. With respect to the greater level of detail
timestamps may also be important in modelling time out behaviour of heartbeats.
The inclusion of time may also allow us to model the retention of keys before
the deletion of pseudonyms. TC's could also consider storing the last $k$
pseudonyms and the analysis would need to ensure that the
adversary could not evade revocation by changing pseudonym at least $k$ times.

Another consideration in an implementation is the handling of cases where no confirmation is sent. If heartbeats are not used then further revocation requests will need to be sent until confirmation is received. 

In the analysis, we currently focus on functional correctness and authentication.
In future work we will consider generalising the correctness analysis, in particular G5, to include liveness properties such that we could prove a more general property such as ``any revocation request must eventually be confirmed''. The \tamarin tool chain has been extended in a recent paper by Backes {\it et al.}~\cite{backes2017} to enable verification of liveness properties.
Not considered here are privacy requirements such as unlinkability which could likewise merit a formal analysis.

Delaune \& Hirschi~\cite{Delaune2017127} and Chadha et al.~\cite{DBLP:journals/tocl/ChadhaCCK16}
survey various anonymity and privacy related properties, including anonymity,
unlinkability and strong secrecy, which can be proved using equivalence-based
reasoning. Behavioural equivalence allows us to determine whether two situations
are different, in particular whether the confirmation of a revocation came from
one vehicle or another.  The use of process equivalences to analyse privacy properties can also be seen in
\tamarin~\cite{DBLP:conf/ccs/BasinDS15} and in other modelling tools, e.g.
\proverif~\cite{blanchet2015proverif}, which has been used successfully to
analyse privacy properties~\cite{fazouane:2015:formal-verifica,DBLP:conf/esorics/DahlDS10}. Future work will be to explore anonymity and privacy
properties of revocation protocols and of other V2X protocols.

Our proposed protocol requires a trusted device at the car which can be trusted to erase all of
  the pseudonyms that the car may have available. 
 However, it is still under debate whether this is the right trust model for the car.  Furthermore, which functions is it reasonable to place within this trusted device, and which cannot be made trustworthy?  To answer these questions in a satisfactory way is not straightforward, and to make the vehicle industry reach agreement on a specific trust model is even more demanding.  This is an interesting challenge for future work.


{\bf Acknowledgements.} Jorden Whitefield is funded by EPSRC iCASE studentship
15220193 through Thales UK. Thanks to Cas Cremers for detailed discussions on \tamarin.  Thanks
also to Fran\c{c}ois Dupressoir and Adrian Waller for detailed feedback, and to the reviewers for their constructive comments.

%
%
\bibliographystyle{splncs03}
\bibliography{bibliography}

\end{document}